\magnification=1200
\baselineskip=18truept
\input epsf

\def\preprint{Y}
\def\draftversion{N}
\def\cap{\hsize=4.5in}

\if \draftversion Y


\fi

\def\cap{\hsize=4.6in}
\def\figure#1#2#3{\if \preprint Y \midinsert \epsfxsize=#3truein
\centerline{\epsffile{figure_#1_eps}} \halign{##\hfill\quad
&\vtop{\parindent=0pt \hsize=4.6in \strut## \strut}\cr {\bf Figure
#1}&#2 \cr} \endinsert \fi}

\def\captionone{\cap
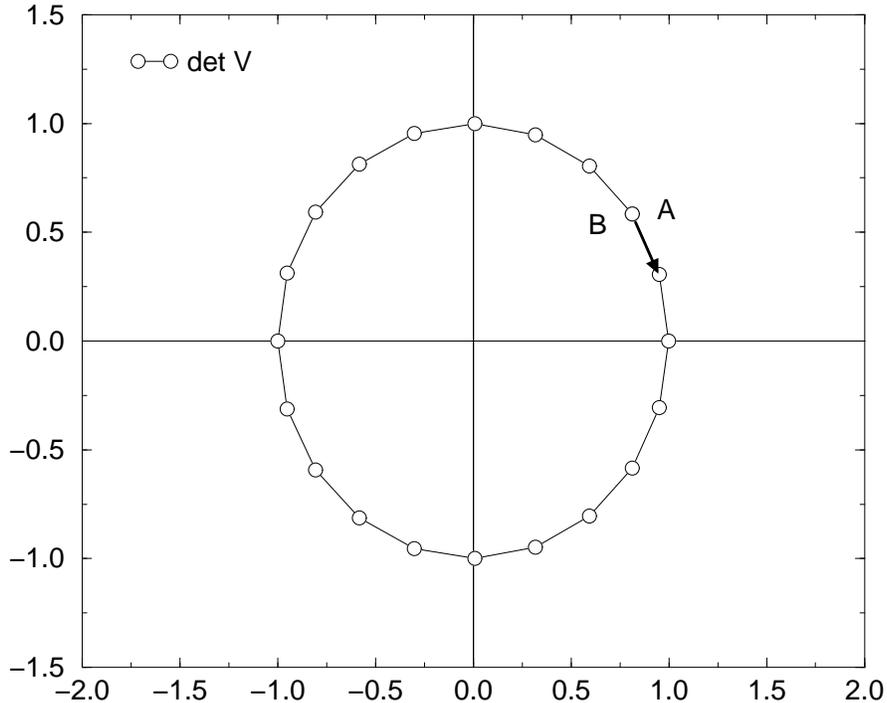
Path of $\det (V)$ in the complex plane as $\theta$ 
varies from $0$ to $2\pi$ at $a=-{{ \pi}\over 5}$. A and B are 
gauge equivalent and the beginning and end of the path.
Points along the path are connected to guide the eye and are
equally spaced in $\theta$. The arrow emanating from A indicates
the sense in which the path is traversed for decreasing $\theta$.}
\def\captiontwo{\cap
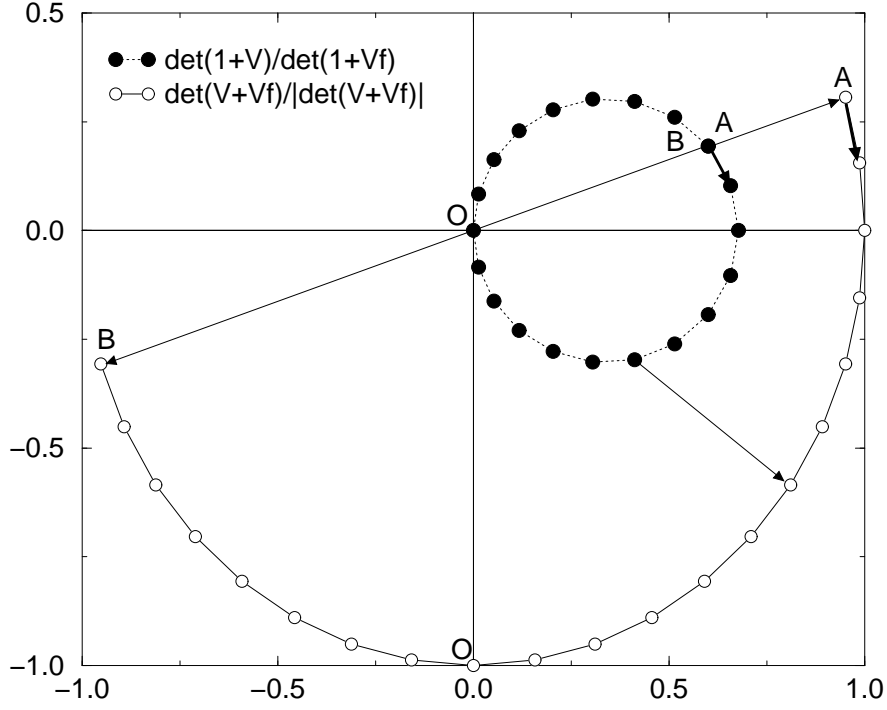
Path of ${{\det (1+V) }\over {\det (1+V_f ) }} $ 
and of ${{\det (V+V_f )} \over {|\det (V + V_f )|}}$ 
in the complex plane for the
same set of configurations as in Figure 1. 
We see the single zero crossing at O by the gauge 
invariant factor (the circle).
The Brillouin-Wigner phase factor is a semi-circle making the 
weights of configurations
A and B differ by a sign. Note that the shape 
of  $\det ({1\over 2} (1+V) )$ 
looks close to being $\propto (1+e^{i\theta})$, 
the one dimensional answer.
Points along the paths are connected to 
guide the eye. The arrow emanating from A indicates
the sense in which the path is traversed for decreasing $\theta$. 
A few corresponding
points on the two paths are connected by lines ending in arrows. }

\def\npblong{1}
\def\plbfirst{2}
\def\narnish{3}
\def\kimura{4}
\def\redlich{5}
\def\alvarezgaumemoore{6}
\def\costeluscher{7}
\def\ho{8}
\def\niemiseme{9}
\def\witt{10}
\def\elitzur{11}
\def\klinghammer{12}
\def\nish{13}
\def\kapsch{14}
\def\kap{15}
\def\anomnn{16}
\def\anomDS{17}
\def\phasechoiceDS{18}
\def\curcivene{19}
\def\huetnn{20}
\def\schwartz{21}
\def\jackreb{22}
\def\dunne{23}
\def\deser{24}
\def\fosco{25}
\def\holgerintanom{26}
\def\minfty{27}
\def\412{28}
\def\num412{29}
\def\polyakov{30}
\def\hsu{31}
\def\nnprl{32}
\def\nnu1{33}
\def\fivedim{34}

\line{\hfill RU--97--61}
\line{\hfill KUNS 1458 HE(TH) 97/12}
\vskip 2truecm
\centerline{\bf Overlap in odd dimensions.}

\vskip 1truecm

\vskip .1in
\centerline{Yoshio Kikukawa\footnote{*} {Permanent Address: 
Department of Physics, Kyoto University, 
Kyoto 606-01, Japan.} and Herbert Neuberger}
\centerline{\it Department of Physics and Astronomy}
\centerline{\it Rutgers University, Piscataway, NJ 08855-0849. }
\vfill
\centerline{\bf Abstract}
\vskip 0.75truecm
In odd dimensions the lattice overlap formalism is
simpler than in even dimensions. Masslessness of
fermions can still be preserved without fine tuning and 
gauge invariance without gauge averaging can
be maintained, although, sometimes, only at the expense
of parity invariance. When parity
invariance is enforced invariance under small
gauge transformations can be maintained and 
continuum global gauge anomalies are reproduced. 

\vfill
\eject
\centerline{\bf 1. Introduction.}
\medskip
In even Euclidean dimensions, 
the overlap formalism [\npblong ]
provides a way to put Weyl fermions interacting with gauge
fields on the lattice. Based on an internal quantum
mechanical supersymmetry [\plbfirst],  it has a mechanism
of protecting masslessness
of fermions. In this paper we deal with field theories in
odd dimensional Euclidean space 
where chirality is absent. Still, protecting masslessness
by the regularization is of potential interest and we are employing
the overlap to do this. These issues were recently addressed in
ref. [\narnish]. 
\smallskip
\leftline{\sl 1a. Continuum features.}
\smallskip
Before we describe our objectives more specifically, let us briefly
recall some basic continuum features. Unless stated otherwise, 
we work on a compact
torus to avoid infrared issues.  
Consider, for example,
a single massless Dirac $SU(2)$ doublet interacting with $SU(2)$ gauge fields
in three Euclidean dimensions:
$$
{\cal L} = {1\over {2g^2}} Tr [F_{\mu\nu} F_{\mu\nu}] - 
\bar\psi \sigma_\mu (\partial_\mu + i A_\mu ) \psi .
\eqno{(1.1)}$$
The $\sigma_\mu$ are Pauli matrices and $SU(2)$ and spinor
indices are suppressed.  
Classically, the Lagrangian possesses
both local gauge invariance and a global discrete symmetry
usually referred to as ``parity''. There also is
fermion number conservation but this symmetry can
be preserved by quantization and we shall ignore it henceforth.
Under parity
$A_\mu (x) \rightarrow -A_\mu (-x)$ and $\bar\psi (x) , \psi (x)
\rightarrow i\bar\psi (-x), i\psi (-x)$. 
A $\bar\psi \psi$ mass term for the fermions 
would switch sign under this parity. It is thought that
not all classical symmetries can be preserved at the quantum 
level:\footnote{${}^{f1}$}{This view may be too 
dogmatic: see [\narnish, \kimura].}
If gauge invariance is kept, parity is broken and, if 
parity invariance is enforced, gauge invariance can be 
maintained only perturbatively, while 
``large'' gauge transformations become anomalous 
[\redlich, \alvarezgaumemoore,
\costeluscher, \ho, \niemiseme].
The global gauge anomaly the theory is afflicted with  
is analogous to the Witten anomaly in four
dimensions [\witt, \elitzur, \klinghammer]. 

Examples for a global gauge anomaly can be found also
for compact $U(1)$ with a unit charge Dirac fermion. 
In the $SU(2)$ case every gauge configuration can be paired
with a gauge equivalent one in such a way  
that it is impossible to maintain
equal weight for both. In the $U(1)$ case only some specific
gauge orbits are anomalous, and if we consider 
non-compact $U(1)$ these orbits disappear. Thus, while there is
an argument that 
parity invariant $SU(2)$ with a single doublet
doesn't exist, this argument does not fully extend 
to $U(1)$ with a single Dirac fermion.

If we replace the
Dirac doublet in the $SU(2)$ case by a Dirac fermion in 
representation $j$ there will be a clash between parity 
and gauge invariance
only when $I\equiv {2\over 3} j(j+1)(2j+1)$ is odd: 
$j={1\over 2}, j={5\over 2}, j={9\over 2}, \cdots$.
When $I$ 
is even a parity invariant and gauge invariant regularization
is possible. Note that $j={3\over 2}$ falls into the category
of ``good'' theories, so it is not only pseudoreality that matters. 
Also, for integral $j$, the representation
of the fermions is real and $I$ is divisible
by 4. Actually, for real representations one can consider
Majorana fermions and then the role of $I$ is played by ${I\over 2}$.
Still, since ${I\over 2}$ is also even there are no problems with
parity for Majorana fermions either. For Majorana fermions with $j=1$ 
the continuum theory also has ${\cal N} = 1$ three dimensional 
supersymmetry. In the Majorana case fermion number is conserved
only modulo 2 and the parity breaking mass terms are of Majorana
type. Unlike in four dimensions, masslessness is not essential
for supersymmetry. However, in conventional lattice approaches
one would need to fine tune to make the gluon and gluino masses
equal. Guaranteeing masslessness is one way to eliminate this 
fine tuning. 
\smallskip
\leftline{\sl 1b. Objective of paper.}
\smallskip
Our main motivation is to see if and how the overlap reproduces
the continuum features just reviewed. 
We would expect the overlap to fully capture at the
regularized level some of the representation dependent 
differences, but the more
subtle distinctions could become apparent only in the continuum
limit. In other words, we wish to subject the basic overlap idea
to a test in odd dimensions and see how close to the continuum
it gets. It is also possible to envisage applications in three
dimensions that employ the overlap [\narnish ,\nish] to other
ends. As a test of the overlap, we are particularly 
interested in the following:
 
An example of global gauge anomaly has not been yet explicitly
exhibited within the overlap formalism. There exist indirect
arguments in favor of the 
overlap reproducing global gauge anomalies:
In [\kapsch] the four dimensional 
Witten anomaly is shown to be reproduced in the continuum version
of the domain wall approach [\kap ] and this indicates that
the lattice overlap will reproduce it also. Also, from
[\elitzur ,\klinghammer ] it is known that the Witten anomaly follows
from the ordinary perturbative anomaly for bigger groups. That
the perturbative anomaly is reproduced by the overlap is quite well
established [\anomnn ,\anomDS ]; thus one should expect 
Witten's anomaly to work correctly also. 
In this paper we look at
the three dimensional analogue of the four dimensional
Witten anomaly and wish to see 
directly how it
gets reproduced on the lattice as a result of insisting
on parity invariance. 

One of the heaviest prices the overlap pays in
even dimensions for dealing with Weyl fermions is the need 
to make a phase choice which 
breaks gauge invariance in fixed bosonic backgrounds. 
There exists a good phase choice and 
although it obeys many desired 
symmetries [\npblong],
it is neither unique (for example see [\phasechoiceDS]), nor 
completely natural. Gauge breaking must be allowed in general 
since for certain representations 
inevitable perturbative anomalies are present. In odd dimensions,
there are no perturbative anomalies  and full gauge invariance
can be preserved (as we saw above,
sometimes at the expense of parity conservation).
We wish to check whether the overlap is sufficiently flexible to
reflect this by providing a more 
natural phase choice than available in even dimensions, 
namely a phase choice which restores full gauge invariance.   
We expect such a phase choice to exist since
preserving full gauge invariance at the expense
of parity is possible using more conventional lattice formulations
than the overlap [\costeluscher]. With the overlap we 
would like to see that also
parity can be maintained in the subset 
of cases indicated by the continuum. 
To achieve parity and gauge invariance
using conventional methods would require fine tuning. 

The simplest
example would be $SU(2)$ gauge theory with a triplet of Dirac fermions,
but a more interesting one would be the case were the fermions
are Majorana and the theory should become supersymmetric in the continuum. 
This would be an analogue of
four dimensions where masslessness is essential 
and where the overlap could also be used to avoid
fine tuning [\curcivene], as pointed out in [\npblong].
To go from Dirac to Majorana in three dimensions should be just
as easy as going from Weyl to Majorana-Weyl in 2 and 10 dimensions
[\huetnn]. In other words, the fact that for real representations
the Dirac theory admits a ``perfect square root'' in the continuum
is something we expect the overlap to reproduce exactly on the lattice.
As suggested in [\huetnn], employing dimensional
reduction from $d=10$
is a possible way to obtain lattice versions of other would be 
supersymmetric theories. Similarly  [\nish], starting from
other dimensions and using the appropriate fermions, more
examples exist [\schwartz].
\smallskip
\leftline{\sl 1c. Synopsis of results.}
\smallskip

In [\narnish ] it was noted that using the same phase choices as
in even dimensions, the so called ``Brillouin-Wigner'' convention,
one could maintain exact parity invariance on the lattice
and gauge invariance under ``small'' gauge transformations. 
Like in four dimensions one needs to rely
on hopefully inoffensive gauge averaging to restore full gauge
invariance in the continuum limit. 
The new points we make here are:
\item {$\bullet$} We present examples of 
global gauge anomalies on the lattice where expected
by the conventional, continuum arguments. 
We identify the underlying lattice mechanism causing these
anomalies. Thus, we see no compelling evidence
for the new universality classes, beyond those of [\costeluscher],
suggested in [\narnish]. To us it 
seems more likely that gauge averaging will eliminate any
interesting continuum limit whenever global anomalies are
thought to exist in the continuum 
on all gauge orbits, like for $SU(2)$
with one doublet. For $U(1)$ with a single Dirac fermion,
the simplest possibility is that gauge averaging eliminates
some of the gauge orbits in the continuum limit, but leaves
an interacting theory which at infinite volume is indistinguishable
from the non-compact case.    
\item {$\bullet$} We show that the overlap formalism simplifies
significantly in odd dimensions allowing for gauge invariant phase
choices. This simplification opens the door to the application
of numerical methods to the study of those non-perturbative phenomena
in three dimensions where massless fermions are essential. In some
cases the computational cost seems not too daunting. The simplification
we find has to do with the fact that in odd dimensions we typically
don't have
to worry about fermion number non-conserving processes and therefore
there is no {\sl a priori} reason to prohibit discretizations 
of the relevant Dirac operators by finite matrices of fixed shape.
In even dimensions the need to accommodate dynamically changing
matrix shapes prohibits the level of explicitness achievable here.
When Higgs fields in a coset space of the gauge group divided
by a subgroup containing 
abelian factors are introduced the story changes
since robust fermion zero modes can occur in three dimensions 
[\jackreb] in the infinite volume limit.\footnote{${}^{f2}$}{
The index of any compact elliptic differential operator on an odd 
dimensional manifold is known to be zero. For nontrivial
indices we need an open space in odd dimensions.}
The introduction of the Higgs fields in the overlap
also spoils the applicability of the simplification,
bringing us back to a formalism where fermion number
violation cannot be ruled out {\sl a priori}.

The plan of the paper is as follows: In section 2 we introduce the
basic (simplified) formulae describing the overlap for
Dirac and Majorana fermions in odd dimensions. In section 3 we present
the simplest possible odd dimensional model: it has $d=1$ and a 
$U(1)$ gauge group. Although this quantum mechanical model
is trivial in itself, it contains mathematical
ingredients that teach us useful 
lessons about higher odd dimensions, 
as has been noted recently in [\dunne, \deser, \fosco]. 
In odd dimensions the $d=1$ model plays an illustrative role similar to
the one played by $d=2$ models [\holgerintanom, \412] in even dimensions. 
In section four we
discuss the global anomaly in $d=3$ 
dimensional models with gauge groups $U(1)$ and $SU(2)$. 
We end with our main conclusion that the
overlap appears to work well in odd dimensions, and that further work
might yield new results.  

\bigskip
\centerline{\bf 2. Basic formulae.}
\medskip
\smallskip
\leftline{\sl 2a. Even to odd dimensional reduction.}
\smallskip
Rather than follow [\npblong ] and derive the odd dimensional overlap
from scratch [\narnish ], it is simpler to view the (odd)
$d$-dimensional theory as a dimensionally reduced  theory
descending from one dimension higher, where the overlap has been
worked out already. In the higher dimension one direction is shrunk 
to a circle of vanishing radius. We only need to worry about
the fermions in a fixed gauge background, so our dimensional
reduction is quite trivial in the sense that it deals 
with a super-renormalizable bilinear action. 
The Weyl fermions in $d+1$
dimensions become massless Dirac fermions in $d$ dimensions. 
We set the gauge field component in the shrunk direction to zero
(later we shall set it to another constant and induce a mass term
this way) and the fermion is assumed to obey periodic boundary
conditions in the same direction. We choose the other components
of the gauge field as independent of the shrinking coordinate and
ignore the infinitely heavy Dirac fermions which make up parity doublets
from the $d$ dimensional point of view. 

This is the setup in continuum. On the
lattice we pick a torus with sides consisting of 
$L$ lattice spacings in the extended
directions and one side of minimal length (one lattice 
spacing) in the shrunk direction (all directions
are compactified, but one is minimal in size). 
The link variables on the
self-closing links are set to unity. The Weyl overlap for this
degenerate $d+1$ dimensional 
lattice should provide the overlap formulation for a massless Dirac
fermion in $d$ dimensions. For definiteness we choose $d=3$,
but it is easy to generalize. 
 
In four dimensions the chiral 
determinant is replaced at the regulated level by
the overlap of two many-body states [\npblong]. 
These are the ground states
of two bilinear Hamiltonians,
$$
{\cal H}_4^\pm = a^\dagger H_4^\pm a \eqno{(2.1)}$$
with all indices suppressed. The matrices $H_4^\pm$ are obtained from
$$
H_4(m) = \pmatrix {B_4 +m & C_4 \cr 
C_4^\dagger & -B_4 -m \cr}\eqno{(2.2)}$$
with $H_4^+ = H_4(\infty )$, $H_4^- = H_4(-m_0 )$ and $0 < m_0 < 2$.
The infinite argument for $H_4^+$ can be replaced by any finite positive
number, but the equations are somewhat simpler with our choice
[\minfty, \412, \num412]. The matrices $C_4$ and $B_4$ are given below:
$$
\eqalign{
( C_4 )_{x \alpha i, y \beta j}
& ={1\over 2} \sum_{\mu=1}^{4} \sigma_\mu^{\alpha\beta}
[\delta_{y,x+\hat\mu} (U_\mu (x) )_{ij} -
\delta_{x,y+\hat\mu} (U_\mu^\dagger (y))_{ij}] ,\cr
( B_4 )_{x \alpha i, y \beta j} & = {1\over 2} \delta_{\alpha\beta} 
\sum_{\mu =1}^4 [2\delta_{xy}
\delta_{ij} - \delta_{y,x+\hat\mu} (U_\mu (x) )_{ij} -
\delta_{x,y+\hat\mu} (U_\mu^\dagger (y))_{ij}].\cr} \eqno{(2.3)}$$
$x,y$ are sites on the lattice, $\alpha ,\beta$ are Weyl spinor
indices and $i,j$ are color indices. The $\sigma_\mu$ are Pauli
matrices for $\mu=1,2,3$ and $\sigma_4 =i$. 
Eliminating the dimension $\mu=4$ we obtain
$$
{\cal H}_3^\pm = a^\dagger H_3^\pm a,~~~~~~~H_3 (m) = \pmatrix{
B_3 + m & C_3 \cr C_3^\dagger & -B_3 -m \cr},\eqno{(2.4)}$$
with $C_3$ and $B_3$ given by eq. (2.3), 
only the indices $\mu$ run from 1 to 3
now and $x$ and $y$ label the sites on a three dimensional lattice.
$\alpha ,\beta ,i ,j$ maintain their original ranges. The elimination
of $\sigma_4$ implies $C_3 = -C_3^\dagger$. The overlap is defined as
$<+|->$. The bra 
and ket are, respectively, 
vacua for (2.1): $ {\cal H}_3^\pm |\pm > = E^\pm_{\rm vac} |\pm >$ 
and there is a phase freedom we shall determine later. 
\smallskip
\leftline{\sl 2b. Gauge invariant / parity non-invariant form.}
\smallskip
$H_3 (m)$ can be brought into a more convenient form by 
unitary transformations acting in the block space made explicit in
(2.2). Conjugating by ${1\over \sqrt{2}} 
e^{i{\pi\over 4}} (1-i\sigma_3 )$, 
we obtain:
$$
H_3 (m)\rightarrow \pmatrix {1&0\cr 0& i\cr } 
H_3 (m) \pmatrix {1&0\cr 0& -i\cr }
=
\sigma_3 \otimes (B_3 +m) - i\sigma_1 \otimes C_3 .\eqno{(2.5)}$$
The $\sigma$ matrices in (2.5) 
operate in a different space from the ones
in (2.3). 
We now rotate in the three dimensional 
space associated with the three $\sigma$ matrices of (2.5)
around the unit vector 
${1\over\sqrt{3}} (1,1,1)$  by an angle of 120 degrees via conjugation by 
$U={1\over 2} [1-i(\sigma_1 + \sigma_2 +\sigma_3 )] $ and obtain
(reusing the $H_3$ symbol) 
$$
H_3 (m) = \sigma_1 \otimes (B_3 +m) -i\sigma_2 \otimes C_3 =
\pmatrix {0 & X\cr X^\dagger & 0 },\eqno{(2.6)}$$
with $X=B_3 + m + C_3$.

This form is known to lead to an explicit formula for the overlap
in terms of the determinant of some finite, fixed dimension matrices
[\npblong]. Making a {\sl natural} phase choice one obtains:
$$
<+|->_{D_c} = \det ({1\over 2} ( 1 +V))~~~
~~~V\equiv X{1\over\sqrt{X^\dagger X}}
~~~~~~V^\dagger V =1 .\eqno{(2.7)}$$
Under a gauge transformation $X$ changes by conjugation so $<+|->_{D_c}$
is fully gauge invariant. Gauge fields where $\det (X) =0$ are assumed
of zero measure and ignored. The subscript $D_c$ 
stands for ``Dirac fermions in
a complex representation''. It is easy to check that in the free
case ${1\over{1+V}}={{1-V^\dagger}\over {V-V^\dagger}}$ has only
the right poles in momentum space, the doublers being canceled by
zeros in the numerator.

Let us now indicate how in perturbation theory,
that is for smooth gauge backgrounds,
a Chern-Simons term with the expected coefficient
is induced in the effective action. This
result can be easily extracted from a 
calculation in [\costeluscher], 
observing that $X$ is precisely the Wilson Dirac lattice
operator with a negative mass. The 
phase of $\det(X)=\det (V)  |\det (X)|$ has been
computed in [\costeluscher] (to leading order and for $0< m_0 <1 $) 
in a context in which
it gives the {\sl infinite}  mass limit answer of the 
continuum.\footnote{${}^{f3}$}{
The conventions we adopted here 
become, in the notation of 
[\costeluscher], $s=1,~m<0,~\gamma_1\gamma_2\gamma_3 = i$.
To use the result of [\costeluscher] first make
the inconsequential overall sign change $D\rightarrow -D$.
Therefore, we need the result of [\costeluscher] 
for $s=-1,~m>0,~\gamma_1\gamma_2\gamma_3 = -i$.
The result of [\costeluscher] is derived assuming 
$\gamma_1\gamma_2\gamma_3 = i$. To account for
the different representation we chose, 
we need to take the complex conjugate of the answer of
[\costeluscher]. 
In the notation of [\costeluscher] we end 
up with $c_0=\pi$. In our next sub-section we 
obtain $c_\infty =0$,
agreeing with the $n=0$ case of 
equations (1.10-11) of [\costeluscher].} 

But, the phase of $<+|->_{D_c}$ in (2.7) is just half the phase
of $\det(V)$ (in perturbation theory)
and this is the correct massless continuum result 
(more precisely, as explained in [\costeluscher], it is one of the
possible correct results).

We shall discuss parity in some detail in sub-section 2f below.
For the time being we just assert that parity invariance would hold
if the expression for the fermion determinant were real for all
gauge backgrounds. 
One cannot find a phase to make $<+|->_{D_c}$ real and rescue
parity invariance while preserving gauge 
invariance because one would then 
need to define $\sqrt{\det (V)}$ for all backgrounds. Although $V$ itself
is sufficiently local, picking a branch of the square root would
either amount to a nonlocal phase choice
or violate gauge invariance. The previous paragraph 
only describes the leading polynomial
part of the imaginary part of the logarithm of $\det({1\over 2} (1+V))$.
While the result of the perturbative
computation is local, 
the numerical value of the coefficient
implies also the existence of a nonlocal term 
if gauge invariance is invoked. 
The nonlocal term takes the discrete values $0$ or $i\pi$ 
depending on the gauge field. 

\smallskip
\leftline{\sl 2c. Massive fermions.}
\smallskip
Let us now introduce a mass
term for the fermions. This is interesting, since in the infinite mass limit
new phenomena may occur [\costeluscher, \polyakov]. As alluded before, 
to add a mass term we again  
dimensionally reduce from four dimensions,
only this time keeping a constant but nonzero fourth component for the vector
potential. On the lattice this produces a shift in our parameter $m_0$, which
we absorb, maintaining the new parameter $m_0$ in the same
range as before ($0< m_0 < 2$), and also a new term in $H_3$. 
Going to the basis used in (2.6) the new Hamiltonian matrix becomes:
$$
H_3^\mu (m) = \pmatrix {\mu & X\cr X^\dagger & -\mu \cr }.\eqno{(2.8)}$$
Note that the fermion mass is given by $\mu >0 $ while $m_0$ is an intrinsic
mass to be interpreted as the mass of extra regulator fields. $m_0$
goes to infinity when measured on physical scales in the continuum limit.
The formula for the overlap in the complex Dirac case (2.7) changes
only in that $V$ is replaced by $V^\mu$, which is no longer unitary: 
$$
V^\mu =X{1\over {\sqrt{X^\dagger X + \mu^2 } + \mu }}.\eqno{(2.9)}$$
If we keep $\mu$ fixed and approach the continuum limit, the mass of
the physical fermion becomes infinite. If we scale $\mu$ to zero
with the gauge coupling constant instead, 
the fermions will acquire a finite mass.
If we take the lattice-$\mu$ to infinity 
the physical fermion is infinitely massive.
The large lattice-$\mu$ limit is obviously smooth. We conclude that
at infinite physical fermion
mass the overlap becomes gauge field independent 
because $V^\mu$ vanishes then.
Thus, no Chern-Simons term is induced, nor are 
any of the special effects
associated with it. Again, we obtained one of the possible continuum
results according to [\costeluscher], albeit the most uninteresting one.

\smallskip
\leftline{\sl 2d. Adjoint  Higgs Fields.}
\smallskip
If the fourth component of the vector potential in the
parent theory is an arbitrary function of the unreduced directions, 
we end up introducing a Higgs field in the adjoint representation. 
On the lattice we would 
write  $U_4 (x_1 , x_2, x_3 )= e^{iH(x_1 , x_2, x_3)}$, 
expand to second order in $H$, and also add an invariant 
potential for $H$ to the rest of the action. Now,
the parameter $\mu$ in (2.8) is replaced by a matrix which
no longer commutes with $X$ because of the non-trivial
$x$ dependence. As a result, the road to the more
explicit expression (2.7) is blocked. This is just as well, 
since now fermion violation might become possible in fixed monopole 
backgrounds (at least in the infinite volume limit)
and an expression for the fermionic path
integral in terms of a finite matrix of fixed shape would at least 
require awkward explanations. The overlap is still
potentially useful, but it has the same
form as in even dimensions, and further simplification
cannot be achieved. We shall not deal with Higgs fields
in the rest of this paper. If we freeze the Higgs field, we go
back to adding a mass term for the fermions, as in sub-section 2c.
\smallskip
\leftline{\sl 2e. Real representations and Majorana fermions.}
\smallskip
Assume the fermions
are in a {\sl real} (not complex and not pseudoreal) representation
of the gauge group. Thus all the $U_\mu (x)$-matrices 
are real. It is easy
to see then that:
$$
\sigma_2 X \sigma_2 = X^T .\eqno{(2.10)}$$
This implies that $\sigma_2 V$ is 
not only unitary but also antisymmetric.
Thus,
a special square root of $\det (V)$ 
exists: $\det (V) = (pf (\sigma_2 V))^2$.
The pfaffian is analytic in the entries of $V$, 
just as the determinant is.
It is also invariant under real gauge transformations.  
We end up with a gauge invariant {\sl and} parity invariant overlap,
but only for real representations carried by fermions of Dirac type.
$$
<+|->_{D_r} = {{[pf ({1\over 2} (\sigma_2 +\sigma_2 V ))]^2}
\over  {pf (\sigma_2 V)}}.\eqno{(2.11)}$$
$D_r$ stands for ``Dirac fermions in a real representation''. 
The main factor in the overlap (representing the gauge invariant but
parity violating version) is also 
a ``perfect'' square, directly implying 
a formula for the case that the representation is real
and the fermions are of Majorana type:
$$
<+|->_M =  pf ({1\over 2} (\sigma_2 +\sigma_2 V )).\eqno{(2.12)}$$
$M$ stands for ``Majorana fermions''. If we want a parity invariant
formula we again face the problem of taking a square
root of $pf (\sigma_2 V )$ without violating locality
or gauge invariance. 

In summary, for Dirac fermions in a real representation we have
complete gauge invariance and parity invariance on the lattice.
For Majorana fermions, if we have exact 
parity invariance, gauge invariance 
cannot be fully maintained, although it could be recovered
in the continuum limit by gauge averaging. We might wish to
reverse the roles of gauge invariance and parity invariance, by
preserving gauge invariance and adding an explicit counter-term
to cancel the induced Chern-Simons term in the continuum limit and thus
restore parity in the continuum. 
However, this requires a gauge invariant lattice construction 
of the exponent of the Chern-Simons action. 
\smallskip
\leftline{\sl 2f. Removal of parity breaking.}
\smallskip
Three dimensional ``parity'' is a descendant of four dimensional
parity. We know that in four dimensions [\npblong] 
replacing $\{ U_\mu (x) \} $ by a parity 
image induces $H_4 (m) \rightarrow H_4^\prime (m) = 
-\sigma_2\otimes 1~ H_4(m) ~ \sigma_2\otimes 1$. 
This implies $H_3 (m) \rightarrow H_3^\prime (m)= H_3(m) 
{|}_{C_3 \rightarrow -C_3}$
or $V\rightarrow V^\prime = V^\dagger$.
Therefore, parity is broken by the overlap
being non-real. Although $V$ is not strictly local
there is no fundamental problem since $X$ has a mass term. We have 
obtained a formulation quite similar to the conventional one
[\costeluscher], only that the fermions from the point of view
of the overlap are massless and are expected to stay so even after
the gauge dynamics is turned on. This expectation is based on an
interpretation involving an infinite number of fermions [\npblong, 
\narnish] and ought to be checked more directly. 

Parity breaking is measured by the ratio
$$
{{<+|->_{D_c} }\over {<+|->_{D_c}^* }} = \det {V} .\eqno{(2.13)}$$
As already mentioned, taking square roots of
quantities with no natural branch
choices could amount to either loss of gauge invariance or
to a nonlocal phase choice.  
We now analyze the issue of ``forbidden'' square roots in the
two problematic 
cases, Dirac with a complex  representation and Majorana, 
in some greater detail. The main observation is that in both cases 
one can define the square of the desired object without violating
gauge invariance, parity and locality (in the generalized sense).
We append a prime to the overlap subscripts to indicate that the
objects themselves aren't yet defined, only their squares are, but
the squares have the nice properties listed above:  
$$
(<+|->_{D_c^\prime} )^2 =
{{(\det ({{1+V}\over 2}))^2}\over {\det (V)}} = 
|\det ({{1+V}\over 2})|^2 =
\det ({1\over 2} ( 1+ {{V+V^\dagger } \over 2})) ,
\eqno{(2.14)}$$
$$
(<+|->_{M^\prime} )^2 = {{\det ({1\over 2} (\sigma_2 +\sigma_2 V ) )}\over
{pf (\sigma_2 V )}}. \eqno{(2.15)}$$

In each case we deal with gauge invariant and real expressions. Gauge
invariance implies that in a complete theory the
gauge action will control the fluctuations of the squares. As the
lattice $\beta$-gauge coupling is taken to 
infinity the behavior of the above
squares would be determinable from the continuum theory. 
It is then 
essentially a question about the continuum whether the above
squares have gauge invariant square roots or not. (We needn't worry
about renormalization effects, as we are really focusing
only on the imaginary part of the effective action.) For $SU(2)$ with Dirac
fermions and $j={1\over 2}$ we expect problems, but for the Majorana
case with $j=1$ we don't. What kind of problems can occur ? 
Clearly, the right hand side of
(2.14) is non-negative. So, all that can go wrong with
preserving the gauge invariance of the square root
is that the square have odd rank zeros as we smoothly deform
one link configuration into a gauge equivalent one. 
This is exactly how global gauge anomalies occur. 

Based on the continuum, we
expect to be able to safely take the square root
(up to gauge configurations of vanishing weight) in the Majorana
case, but we suspect that it is impossible to 
carry out this step in the Dirac case. 
If we are right about the Majorana case we obtain a quite elegant
lattice regularization for supersymmetric YM in three 
dimensions. This has an analogue
in four dimensions [\npblong], where one could use the absolute value
of the 
appropriate overlap, again relying on the continuum for 
locality, in particular
on the recent claim that the gluino 
determinant has no odd rank zeros [\hsu]. 

Some indirect evidence in favor of such a procedure can be
found in the recent work testing the overlap in two dimensions [\num412].
There, a particular chiral model was investigated which is
known to have a positive definite fermion determinant (when all species
are combined) in the continuum. It was found numerically that
carrying out gauge averaging (and thus respecting locality) gave
continuum results indistinguishable within errors from a
nonlocal procedure where gauge invariance was enforced by taking
the absolute value of the overlap. Apparently, if there is no
conflict between the nonlocal expression and the continuum limit, 
the lattice nonlocality leaves no traces in the continuum limit.
\smallskip
\leftline{\sl 2g. Parity invariant / 
gauge non-invariant Brillouin-Wigner phase choice.}
\smallskip  
We now consider the Brillouin-Wigner
phase choice. With this phase choice a more direct
comparison with [\narnish] can be made and this
is the phase choice we employ in even dimensions, so it
is interesting to see how well it does here. For brevity, 
we discuss only
a Dirac fermion in a complex representation. 
Let us first deal with the massless case, $\mu=0$, since the main
issue is reality, which is anyhow absent if $\mu\ne 0$.
The Brillouin-Wigner
convention amounts to fixing the phases of the $|\pm >$ 
states by requiring
they have positive overlaps with corresponding states in the
absence of nontrivial gauge fields. We shall denote the appropriate
matrices by the subscript $f$ for ``free''. It is quite easy to see
that:
$$
<+|->_{D_c}^{\rm BW} =  \det ({1\over 2} ( 1 +V)) z^{\rm BW} ,~~~~~~~
z^{\rm BW}= {{\det ({1\over 2} ( 1 +V^\dagger V_f 
))}\over{| \det ({1\over 2} ( 1 +V^\dagger V_f ))|}} { 
1\over{\sqrt {\det ( V_f ) }}} .
\eqno{(2.16)}$$
The constant phase factor ${1\over
{ \sqrt {\det ( V_f ) }}} $ is added for convenience. Actually, it is not hard
to see that $\det (V_f ) =1$ for reasonable fermion boundary conditions.
There is no gauge invariance and the expression is local in the
sense that it involves no dangerous square roots. To see that the
expression is real we compute its square:
$$
(<+|->_{D_c}^{\rm BW})^2 = |\det ({1\over 2} (1 +V)) |^2 .
\eqno{(2.17)}$$
We find the same answer as in (2.14) and conclude 
that $<+|->_{D_c}^{\rm BW}$ is one possible candidate
for $<+|->_{D_c^\prime}$ and, since (2.16) is 
essentially local, it had to violate gauge invariance.
The gauge violation is made explicit by the presence of 
the matrix $V_f$, 
which does not rotate under gauge transformations. However,
the gauge violation is more minimal than it appears at 
first sight: equation (2.17) can be 
inverted to write $( z^{\rm BW} )^2 =\det^* (V)$
showing that $( z^{\rm BW} )^2$ is gauge invariant. 
For $V=V_f$, $z^{\rm BW}={1\over\sqrt{\det (V_f )}} ~(=1)$
and by continuity we see that there is gauge invariance for
``small'' gauge transformations. 
If the Brillouin-Wigner phase were completely well defined 
for any gauge background there would be no room for a problem.
But it is not: there are exceptional 
configurations for which $VV_f^\dagger$ has $-1$ as an eigenvalue
and $z^{\rm BW}$ becomes ill defined. It is safe to ignore
these configurations in the path integral -- but their presence
thwarts any continuity arguments beyond ``small'' gauge transformations.
So gauge invariance is almost restored, but sign changes can occur
under certain gauge transformations even in the continuum limit. 

That this
is the case will be argued in the next section. These sign changes are
the lattice version of the global gauge anomaly. We have learned that
the Brillouin-Wigner phase choice is indeed a ``good'' one, in the sense
that it goes as far as it only can, given the continuum results 
and moreover, it does not go beyond these, 
thus potentially respecting the inevitability
of global gauge anomalies as long as only
local counter-terms are allowed. 

Consider two gauge equivalent configurations (A and B)
that are known in the continuum to give opposite
signs to the square roots when the latter are defined in
a local way, by using the ``doubling trick'' of
[\witt]. When A is connected to B by a smooth path 
in the continuum, we expect, on the lattice, 
the gauge invariant
and parity non-invariant version (2.7) to have a total
odd number of zeros along the path. We don't 
expect $z^{\rm BW}$ to have to vanish at all along the
path, 
since we should be able to deform the path away from
all the exceptional configurations, unless the
endpoints are exceptional themselves. Thus, we expect (2.16)
to have different signs for A and B. 
On the lattice, unlike in the continuum, 
A and B can be connected by a path contained entirely in a single
gauge orbit. However, this path cannot have a continuous
continuum image, no matter how we deform it inside the 
gauge orbit and, we expect it to be
forced to pass through exceptional lattice configurations an
odd number of times. This time the sign switches
occur due to odd zeros in the numerator of
the $z^{\rm BW}$ factor in (2.16). 

The inevitability of the sign switches can be heuristically
understood as follows: 
There is no problem with gauge invariance as long
as the expression is allowed to be complex since the 
fermion determinant, as a function of the parameter
describing the path connecting A to B, can trace out  
a continuous closed curve in the
complex plane even though it goes
through zero an odd number of times.
But, projecting the curve on the real axis (or any other
smooth line going through the origin once)  
leaves no way for it to close if it crosses zero an odd number of times. 
Under the Brillouin-Winger phase choice gauge invariance is broken
and the image of the curve on the real axis opens up, leaving $z^{\rm BW}$
with opposite signs at A and B.

For $\mu\ne 0$ the $V$'s in (2.7) and (2.16) are replaced by
the $V^\mu$ of (2.9). In sub-section 2c we noted that
at infinite $\mu$ $\det (1+V^\mu ) = 1$. 
The phase $z^{\rm BW}$ also 
goes to one in the infinite $\mu$ limit. Therefore, in
the infinite mass limit, no Chern-Simons term is induced
in either one of the two definitions (2.7) and (2.16).
Our $\mu$--mass term in (2.8) is identical to that of [\narnish].
Our choice of $H_4^+ = H_4 (\infty )$ makes our overlap
definition somewhat different from the one 
in [\narnish]. This
difference should be insignificant in the continuum limit.
Nevertheless, 
our conclusion about the infinite mass limit of the continuum
theory differs from [\narnish] and our fermion truly decouples
when it becomes heavy. 

Note that
(2.17) does not hold in the massive case since 
$V^\mu$ is no longer
unitary. 
\vfill\eject

\bigskip
\centerline{\bf 3. A one dimensional example.}
\medskip
\smallskip
\leftline{\sl 3a. Definition and continuum solution.}
\smallskip
The subtle interplay between parity and gauge invariance for 
odd dimensional Dirac operators extends to $d=1$ [\dunne,\deser].
The Lagrangian has no kinetic term for the gauge field and the
fermions have a single spinor component. We take $U(1)$ as the
gauge group since the nontriviality of $\pi_1 (U(1))$ is essential.
$$
{\cal L} =- \bar\psi (x) (\partial_x +i A(x) ) 
\psi (x) \equiv \bar\psi (x)
D(A)  \psi (x) ,
~~~~~\psi (0) = -\psi (l),~ \bar\psi (0) = -\bar\psi (l).
\eqno{(3.1)}$$
The eigenvalues of $D(A)$ are $\lambda_n = 
-ip_n$, $p_n = {{2\pi}\over l } ( n - {1\over 2}) +
{\theta \over l}, n\in Z$ where $\theta =-\int_0^l A(x) dx$ is the
one dimensional Chern-Simons action and also the single
purely bosonic gauge invariant variable in the model. 
To regulate the theory gauge
invariantly, we introduce three Pauli Villars fields. Their masses
and statistics are chosen to cancel the leading and subleading
large $n$ behavior of the $p_n$. All the PV fields have 
action similar to $\psi$, only $D(A)$ 
is replaced by $D(A)+M$, where $M$ is a PV mass term. 
Two of the PV fields
are bosonic with masses $m_{1,2}^b$ and one is fermionic with
mass $m^f = m_1^b +m_2^b$. The regulated determinant of $D$ is
$$
\eqalign{
Z^{\rm reg} (A) =& {\cal N} \prod_{n\in Z} {{p_n (p_n +im^f)}\over
{(p_n +im_1^b )(p_n +im_2^b )}}\cr
= & {\cal N} \prod_{n=0}^\infty \left [(1+{\gamma\over{n+\alpha}})
(1-{\gamma\over{n+\beta}})\right ] \prod_{n=1}^\infty
\left [(1+{\gamma\over{n-\beta}})
(1-{\gamma\over{n-\alpha}})\right ]\cr} 
\eqno{(3.2)}$$
where
$$
\alpha= {\theta\over{2\pi}} -{1\over 2} +i{{m_2^b l }\over {2\pi}},~~~~~~
\beta= {\theta\over{2\pi}} -{1\over 2} +i{{m_1^b l }\over {2\pi}},~~~~~~
\gamma= i{{m_1^b l }\over {2\pi}}.\eqno{(3.3)}$$
The normalization ${\cal N}$ is chosen so that $Z^{\rm reg} (0) =1$. 
The infinite products can be done exactly and one ends up with:
$$
Z^{\rm reg} (A) ={\cal N}
{{\sin (\pi (\beta -\gamma )) \sin (\pi (\alpha +\gamma))}
\over {\sin (\pi\alpha ) \sin (\pi\beta )}}.\eqno{(3.4)}$$
Taking $m_{1,2}^b$ to positive infinity we obtain:
$$
Z^{\rm reg} (A) \to {{1+e^{i\theta}}\over 2} = Z(A).\eqno{(3.5)}$$
This agrees with [\dunne, \deser]. Under parity $\theta\rightarrow -\theta$
and $Z(A)\rightarrow Z^* (A)$. $Z(A)$ is gauge invariant because
it is a periodic function of only the variable  $\theta$. 

Suppose we try to achieve parity invariance by ``adding a counter-term''
to extract the phase of $Z(A)$. The phase is $e^{{i\theta}\over 2}$ and
is not periodic. Thus the real answer we are left 
with, $\cos ({\theta\over 2})$
isn't periodic either. More precisely, the original 
periodicity $\theta\rightarrow\theta+2\pi k$ has been 
reduced to $\theta\rightarrow\theta+4\pi k$. Thus, the square of $Z(A)$
can be made both periodic and parity invariant. Taking the square root
is a problem though. 

One can phrase the problem of taking 
the square root in a language entirely
analogous to the flow argument of [\witt]. Define a new theory whose
path integral is directly the square $Z^2 (A)$ 
by doubling the number
of fermions. This is an example of the doubling trick
mentioned in subsection 2g. 
In the new theory $D(A)$ is replaced by $D_2 (A) =\sigma_1\otimes D$. 
Since $\{D_2 (A) , \sigma_3 \otimes 1\}=0$ 
and $D(A)$ is hermitian 
all eigenvalues of $D_2 (A)$ occur in real
pairs of opposite signs. We
could tentatively try to define $\det (D(A))= \sqrt{\det(D_2 (A))}$ 
as the product of all
positive eigenvalues of $D_2 (A)$. Suppose we do that for the 
configuration $A(x)=0$. We now deform 
this configuration to $A(x)={{2\pi}\over l}$
by $A(x;t)=t{{2\pi}\over l}~~~0\le t \le 1$. 
We wish to show that, as a function
of $t$, one pair of eigenvalues of $D(A(x;t))$ cross 
each other at zero for some $t\in [0,1]$ and thus, if we follow the
eigenvalues, we have a sign switch and end up with
the $\sqrt{\det(D_2 (A))}$ at $A={{2\pi}\over l}$ being the negative of
the $\sqrt{\det(D_2 (A))}$ at $A=0$. 
To establish the crossing we introduce
a two dimensional Dirac 
operator $\hat D = i\sigma_2 \otimes {{\partial}
\over {\partial t}} + D_2 (A(x;t))$.
The operator $\hat D$ has an index since it sees a two dimensional
instanton. Therefore  $\hat D$ has one normalizable (chiral) zero mode.
Such a zero mode can exist only because there was one zero crossing
in the spectrum of $D_2 (A(x;t))$ as $t$ varied. 
Clearly, this argument is too
complex for our simple problem, but it is an argument that generalizes
to higher dimensions, where more direct tools cannot be applied.
The main lesson for us is that the square root problem is indeed the
image of the standard global anomaly as understood from the spectral
flow argument. 

In our simple example the problem is evident by inspection of the exact 
result. $Z(A)$ depends only on $\theta$ and as $\theta$ completes
one cycle in the space of gauge orbits $Z(A)$ describes a circle
in the complex plane centered at 1/2 on the real axis and of radius 1/2.
The important feature of the circle is that it passes through the
origin exactly once as $\theta$ goes over its cycle once. 
As a result, the
phase of $Z(A)$ starts at ${\pi\over 2}$ and ends at $-{\pi\over 2}$.
It is not a smooth image
of the $\theta$ cycle since it doesn't have integral winding. 
When $Z(A)$ goes through the origin the
phase is ill defined and this is how periodicity is lost. Extracting
the phase to obtain a real and hence parity invariant expression 
leaves a result that also violates periodicity: The parity
invariant answer has a global gauge anomaly. 

Actually, the simplest way to state the 
difficulty is that $e^{i\theta}$ winds 
an odd number of times as $\theta$ is taken
from $0$ to $2\pi$ and therefore does not admit a smooth
square root. In the general case, on the lattice, it is 
the phase of $\det (V)$ that winds. 
The phase of $\det (V)$, in turn, 
is just the phase of $\det (X)$. $X$ is
a simple and strictly local operator, the
lattice Wilson--Dirac operator in odd dimensions with
a negative mass term. We conclude that the simplest
signature of the global anomaly would be a closed
path in the space of gauge orbits 
along which the phase of $\det(X)$ winds an odd number of times. 
\smallskip
\leftline{\sl 3b. Lattice overlap regularization.}
\smallskip

We regulate the theory as in section 2, only $\sigma_\mu =1$ 
and $\mu\equiv 1$. The spinorial index disappears. 
Since we work with $U(1)$ there are no
gauge indices either. We need the matrix $X$ of (2.6) from
which we should construct the matrix $V$ of (2.7). 
$$ 
(X(m))_{xy} = (m+B_1 +C_1 )_{xy} = (1+m)\delta_{xy} -\delta_{x, y+\hat 1 }
U^\dagger (y)\eqno{(3.6)}$$
The problem simplifies for $m=-m_0 = -1$. Although the main conclusions
are true for all $0< m_0 < 2$ it is easier to derive them at this
particular value and we keep $m_0 =1$ for the time being. At this value
of $m$, $X$ becomes unitary by itself and is the same as $V$.

The gauge invariant, but parity breaking regulated answer is
$$
Z^{\rm reg} (\theta ) = 
{\cal N} \det ({1\over 2} (1+X(-1))).\eqno{(3.7)}$$
The expression being gauge invariant, we can 
replace $U^\dagger (y)$ in (3.6) by $e^{{i\theta}\over L}$, 
where $-\pi < \theta \le \pi$ is defined 
by $\prod_{x=0}^{L-1}U(x) = e^{-i\theta}$
in analogy with the continuum variable. With antiperiodic boundary
conditions imposed we have
$$
Z^{\rm reg} (\theta ) ={\cal N} 
\prod_{n=0}^{L-1} {{1-z_n e^{{i\theta}\over L}}\over 2},\eqno{(3.8)}$$
where the $z_n$ are all the distinct roots of $z^L + 1=0$. Therefore,
$$
Z^{\rm reg} (\theta ) ={{\cal N}\over 2^L} e^{i\theta} (e^{-i\theta} +1)=
{{1+e^{i\theta}}\over 2}.\eqno{(3.9)}$$
We have reproduced the continuum answer (3.5) exactly, without
needing to take the cutoff $L$ to infinity (${l\over{L}}$ is 
our lattice spacing and should be taken to 
zero at fixed $\theta$ and $l$ ). Had
we chosen $0< (m_0 \ne 1) <2$, there would have been some $L$ dependence
at finite $L$ but the continuum limit is approached very rapidly
as long as $m_0$ is not too close to the endpoints $0$ or $2$.
Our discussion about windings in subsection 3a applies therefore
on the lattice too.

The Brillouin-Wigner answer is just as easy to obtain. We find
$$
z^{\rm BW} = \prod_{x=0}^{L-1} {{1+U(x)}\over {|1+U(x)|}}.
\eqno{(3.10)}$$
Note that everywhere the link variables are written excluding
the phase factor that imposes the antiperiodic boundary conditions. 
The exceptional configurations are obvious now: they occur when any
of the $U(x)=-1$. Suppose $U(x)\ne -1$ for all $x$. Write $U(x)=
e^{i a(x)}$. Avoidance of exceptional configurations implies one
can choose $-\pi < a(x) <\pi$ (sharp inequalities at both
ends) for all $x$ and therefore 
$$
z^{\rm BW} = e^{{i\over 2}\sum_x a(x) }.\eqno{(3.11)}$$
The answer with the Wigner-Brillouin phase convention is, for 
unexceptional configurations,
$$
Z^{\rm reg}_{\rm BW} = {1\over 2} (1+e^{i\theta}) z^{\rm BW} =
\cos \left [ {1\over 2}\sum_x a(x)\right ] ,\eqno{(3.12)}$$
exhibiting the global gauge anomaly.

Take two gauge equivalent configurations: Configuration
A has $U(x) =1$ for all $x$ and configuration B 
has $U(x) = e^{i{{2\pi}\over L}}$. One can connect the
two configurations by a 
smooth path $U(x,t) =  e^{it{{2\pi}\over L}}$, $t\in [0,1]$. 
Assuming $L\ge 3$ we see that all exceptional configurations  are avoided 
and $z^{\rm BW}$ is well defined for all $t$. The 
product $Z^{\rm reg}_{\rm BW} = \cos (\pi t )$ 
is therefore also well defined and
real along the path. It switches sign at $t={1\over 2}$. 
This is the single zero crossing along the path.
The zero occurs in the gauge invariant factor. 

The same two configurations can be connected by another 
path: $a(x,t) = t{{2\pi}\over L}$ for $x=0,1,..L-2$ 
and $a(L-1,t) = t ({{2\pi}\over L}-2\pi)$ while $t\in[0,1]$ as before.
This time the path is on a single gauge orbit 
since $\prod_x U(x,t)$ is $t$-independent. Therefore the
gauge invariant factor in $Z^{\rm reg}_{\rm BW}$ stays constant (and
equal to unity). The link variable set is forced to go through
an exceptional configuration exactly once, at $t_* = {{L-1}\over {2L}}$
where $z^{\rm BW}$ is ill defined. For $t\in[0,t_*)$ $z^{\rm BW}=1$
but for $t\in(t_* , 1]$  $z^{\rm BW}=-1$ producing the sign
change. 

There is an even simpler way to describe the problem. 
Consider the smooth
path again. As the path is traversed $\det (V)=\det (X) $ 
must describe a closed
circle in the complex plane. But
$$
\det (X)=(-)^L \prod_{n=0}^{L-1} (z_n  e^{it{{2\pi}\over L}}) .
\eqno{(3.13)}$$
As $t$ goes from zero to one, the quantities 
$z_n e^{it{{2\pi}\over L}}$, 
the eigenvalues of $V$, move on the unit circle 
such that at $t=1$ we have $z_n$ taking the place 
previously held by $ z_{n+1}$ if $n$ labels them
cyclicly round the circle. Each eigenvalue replaces the one following it.
Therefore, the sum of all the eigenvalue 
displacements is $2\pi$. In other
words the determinant of $V$ wound around once. This is why it
cannot have a smooth square root. Conversely, if we only knew that
the determinant of $V$ winds round the origin once, we could conclude
that some eigenvalue crossed the negative 
real axis an odd number of times.
Indeed, if this were not true the sum of all the eigenvalue motions
could not have amounted to $2\pi$. Any time an eigenvalue of $V$
crosses $-1$, $\det (1+V)$ has a simple zero and we can argue as above
for the Brillouin-Wigner case. 

The simpler argument from the previous paragraph has a significant 
advantage: Now the global anomaly can be seen from the
behavior of $\det (V)$ so
we may as well look at the phase of $\det (X)$, even when $V\ne X$  
since $\det(V)=\det(X)/|\det (X)|$. $X$ is a simple, completely
local operator, and its behavior is relatively easy to determine. 
For example, we see now without any effort why our major
conclusions would not change when $m_0 \ne 1$. We also understand
easily why $m_0$ is restricted to the interval $(0,2)$.

The one dimensional 
example has taught us how we expect things to work out in any
odd dimension. The rest of the paper is devoted to three dimensions.

\bigskip
\centerline{\bf 4. Three dimensions.}
\medskip

\smallskip
\leftline{\sl 4a. A path in orbit space for U(1).}
\smallskip
To detect a global anomaly for $U(1)$ we need two
gauge equivalent configurations which have different
Chern-Simons actions, so are connected by a large
gauge transformation. The large gauge transformations are
those that wind non-trivially as the location is taken
round any of the cycles of the torus. For such a winding
gauge transformation to have the effect of changing the
Chern-Simons action, the flux through the two-torus
made up of the other two directions must be non-vanishing. In
a non-compact formulation with periodic boundary
conditions on the vector potentials, such configurations
will not occur in the continuum limit. On finite
lattices, since the fermions only see compact link variables,
no matter what the pure gauge action is, configurations
of the above type are not ruled out. But, they become
extinct as continuum is approached and one needn't
worry about them in a dynamical context
(for non-compact $U(1)$). For our purpose
here the pure gauge action is immaterial. We only want to see
whether an orbit deemed globally anomalous by continuum
arguments will also exhibit a global anomaly on the lattice.

We pick a simple configuration
[\fosco] with constant field 
strength $F_{12}\ne 0$ and uniform $A_3$. 
The total flux of $F_{12}$ through the two torus
spanned by directions 1 and 2 at fixed $x_3$ is quantized and
we pick the minimal uniform value, $F_{12} = {{2\pi}\over l^2}$. 
As $A_3$
increases by ${{2\pi}\over l}$ the Chern-Simons action changes
by $2\pi $. Writing $A_3 = {{a+\theta}\over l}$, 
we expect the fermion determinant to wind around the
origin once as $\theta$ varies between $0$ and $2\pi$, connecting
the two gauge equivalent configurations $A_3 ={a\over l} $ 
and $A_3 ={{2\pi +a }\over l}$. The gauge invariant meaning
of $a$ is in the phase factor $e^{ia}$, which can be thought of 
as a parameter labeling the different pairs of gauge equivalent
configuration we could consider. 

We put this set of configurations on the 
lattice in a straightforward manner.
We pick $-\pi < a \le \pi$ and set 
the links in the third direction
to $e^{i{a\over L}}$. One can also think of $a$ as a parameter
interpolating between different boundary 
conditions in the third direction.
By convention, $a=0$ corresponds
to anti-periodic boundary conditions in the third direction. 
In addition, we 
implement boundary conditions of our choice in the two other
directions.  
The ``instanton'' field in the 1,2 directions is realized as in
[\nnprl]: The parallel transporters 
around all $\{1,2\}$ plaquettes 
are equal to $e^{i{{2\pi}\over L^2 }}$. 
For simplicity, all sides of the torus have $L$ sites.

\smallskip
\leftline{\sl 4b. Argument for the winding of $\det (V)$.}
\smallskip
On the basis of the previous section 
we wish to argue that as $\theta$ is varied $\det(X)$ 
winds round the origin once. This 
will happen if and only if, as  $\theta$
covers its range, an odd number of $V$-eigenvalues cross
the point $-1$ on the unit circle. In turn, 
this is equivalent to counting
how many real negative eigenvalues of $X$ one finds  
as $\theta$ traces out the loop of orbits. Since $tr (X-3-m)^{2k+1} =0$ 
the eigenvalues
of $X$ are paired in $(\lambda , 6+2 m - \lambda)$ pairs, but this is
dependent on the specifics of our discretization and we shall not make
use of it. Clearly, $a$ can be absorbed in $\theta$, so 
we ignore $a$ in what follows and set the boundary 
condition to antiperiodic in all directions
for definiteness. 

Our lattice configuration has translational invariance in the $x_3$
direction so we go to momentum space (labeled by $p_3$) in that 
direction. Because of the antiperiodic boundary conditions $p_3$
takes the values ${{2\pi}\over L} (n -{1\over 2})$ with $n=0,..,L-1$.
$X$ is block-diagonal and the block labeled by $p_3$ is given by:
$$
X_3 (p_3) = 1-m_0 -\cos(p_3 + {\theta\over L}) 
+i\sigma_3 \sin (p_3 + {\theta\over L}) +B_2 +C_2 .\eqno{(4.1)}$$
$B_2$ and $C_2$ are the same as in (2.3) only $\mu$ is 
restricted to 1,2 now.
We are looking for real eigenvalues of $X_3 (p_3)$ for 
any $\theta \in (0,2\pi ]$ and any $p_3$ defined by an 
$0\le n \le L-1$.
If $X_3 (p_3 ) \psi = \lambda \psi$ with 
real $\lambda$, $\psi^\dagger \sigma_3
X_3 (p_3) \psi$ must be real and hence
$$
\sin (p_3 + {\theta\over L})=0~.\eqno{(4.2)}$$
Both $p_3$ and $\theta$ are fixed by this equation to 
either ${\theta\over L } =-p_3 = {\pi \over L}$ 
or ${\theta \over L }= \pi -p_3  
={{2\pi}\over L} ({{L+1}\over 2}-[{{L+1}\over 2}])$ 
where $[x]$ denotes the largest integer smaller or equal to $x$. 
In those cases we have 
$$
X_3 (p_3) = 1-m_0 - \epsilon + B_2 +C_2 \eqno{(4.3)}$$
with $\epsilon = 1$ in the first case and $\epsilon =-1$ in the second.
We are searching for a 
negative eigenvalue $\lambda$ of $X_3 (p_3 ) $.
Define $m_2 =  1-m_0 - \epsilon-\lambda$. We are looking simultaneously
for solutions $\psi$ (with $\psi^\dagger \psi > 0 $ )
and negative numbers $\lambda$ such that 
$$
\sigma_3 ( X_3 (p_3) -\lambda) \psi = H_2 (m_2 ) \psi =0~.\eqno{(4.4)}$$
Here $H_2(m_2 )$ is given by an 
expression similar to (2.2), only in two
dimensions, and the gauge field 
is that of a two dimensional instanton. 
For $\epsilon =-1$, $m_2 > 0$, 
and we know from [\npblong, \nnprl ] 
that $H_2 (m_2)$ has a gap and no zero energy 
solutions. The single remaining
possibility is $\epsilon = 1$. To obtain $\lambda = -m_0 -m_2 < 0$, 
taking into account that $0< m_0 < 2$ and that $m_2 < 0$, we
can search only in the window $-2 < m_2 < 0 $. But,
precisely in this range,
we know, 
again from previous work [\npblong],
that there does exist a unique $ -2 < m_2^z < 0$, which approaches 
zero as $L$ is taken to infinity, where $H_2 (m_2)$ 
has a zero energy eigenstate. As function of $-2< m_2<0$ 
there is exactly one crossing of
zero in the spectral flow of $H_2 (m_2 )$. 
Let the state of zero energy
be denoted by $\psi^z$.  For $L$ large enough that $m_0 >- m_2^z$ 
we have 
$$
X_3 (p_3) \psi^z = \lambda^z \psi^z \eqno{(4.5)}$$
with $ \lambda^z = -m_2^z- m_0 < 0$. 

Our search was exhaustive. We learn that as a function of $\theta$
one eigenvalue of $X$ will cross the negative real axis once.
Thus, $\det (V)$ will circle the origin once as 
theta is varied over its
range and there is no smooth square 
root and there is a global gauge anomaly.

Our argument relied on two dimensional properties that were established
partly numerically. Since the argument is a bit involved, we have carried
out a direct numerical check. Moreover, setting up the numerical check is
worthwhile because we also wish to check explicitly that indeed the
Brillouin-Wigner definition reproduces the global anomaly. For 
the gauge non-invariant -- parity invariant case 
the above does not constitute a full proof since it could be 
that $z^{\rm BW}$ changes sign an odd number of times
as $\theta$ is varied, thus annulling the gauge invariant effect 
of  $\det ({1\over 2} (1+V))$ 
going through its unique zero.

Note that the above argument about the phase of $\det (X) $ essentially
gives the coefficient of the induced Chern-Simons action as computed
in [\costeluscher] perturbatively. We have determined the $\theta$
dependence of $\det (X)$ by picking a set of gauge configurations for
which the three dimensional Chern-Simons action 
becomes identical to the one dimensional Chern-Simons
action. 

If we invert the argument, we could say that
the perturbative computation in [\costeluscher] of the coefficient
of the Chern-Simons term is an indication that in two dimensions
the spectral flow should have the properties desired to
reproduce instanton effects on the lattice. Thus, not only does the
induced Chern-Simons term indicate that the exact anomalies will be
found in one dimension lower [\kap], but it also provides evidence
that instantons should work correctly, 
which is already well known [\npblong].

Also note that had we taken $m>0$ we would have 
found no solutions with $\lambda^z <0$ for $L$ large enough.


\smallskip
\leftline{\sl 4c. Numerical result for U(1).}
\smallskip
\figure{1}{\captionone}{5.5}
\figure{2}{\captiontwo}{5.5}
We checked on the computer that indeed the phase of $\det(V)$ winds round the
origin as argued above. Figure 1 shows the path of $\det (V)$ in the
complex plane for $L=8, m_0 =.8$.

Each value of $a$ defines one pair of gauge 
equivalent configurations $A$ and $B$ where $B$ has $a$ shifted by $2\pi$.
As $\theta$ is varied over one period, $A$ and $B$ 
are connected by a path of gauge in-equivalent intermediary configurations.
We expect $\det (1+V)$ to go through zero once.
We pick $a=-{{\pi}\over 5}$, $L=8$ and $m_0 =.8 $ and trace the 
evolution of $\det (1+V)$ in Figure 2. We see the expected zero crossing 
and that the dependence on $\theta$ is similar
to what we found in the one dimensional model. 
We also follow the $z^{\rm BW}$ phase along
the path. It describes a smooth semi-circle, playing the role
of a square root of $\det (V)$. The configurations $A$ and $B$
have weights of opposite sign therefore. 

However, the picture above is not reproduced for all $a$. As $L$
gets larger it seems that the global anomaly is seen on the lattice
for all $a\in (-2\pi , 0)$, but for $a\in (0, 2\pi ) $ the path
goes through an exceptional configuration, and the overall
sign change cancels, leaving us with identical weights
for the relevant configurations $A$ and $B$. The exceptional
configuration occurs at $a=2\pi$. There might
be a simple explanation for this configuration, but we have
not found one yet. By changing the boundary conditions of the
fermions to periodic in all three 
direction, the range of $a$ (resetting the point $a=0$ to
the antiperiodic case, so we can compare) where
the global anomaly is seen moves to $a\in (-\pi , \pi )$. 
However, exceptional configurations appear now at other values of $a$.
Therefore, it appears that the
Brillouin-Wigner phase choice, for the set of 
configurations we considered here, does not behave
exactly as a continuum definition obtained by using
the doubling trick would.\footnote{${}^{f4}$}
{There is some similarity between this phenomenon
and specific singular gauge transformations in 
two dimensional $U(1)$ gauge theories. In two dimensions
the effect of these gauge transformations
cannot be ignored [\nnu1].} Of course, 
our configurations are quite special.

There is little point to try to quantify
how often such cancelations will happen for more general configurations
since we do not expect global anomalies to play such 
a central role for $U(1)$ in the continuum anyhow. 
The important lesson from the above examples
is the possibility to reproduce global gauge anomalies
for some configurations that are smooth and therefore cannot be ignored
in the continuum limit. Also, one should keep in mind that 
it is not necessary for the Brillouin-Wigner phase convention to
become identical to the definition obtained from the doubling trick in
the continuum. It could be that the doubling trick minimizes in some sense
the amount of gauge breaking while the  Brillouin-Wigner phase convention
does not. This does not deter from the validity of 
the Brillouin-Wigner phase convention when there are no anomalies present.
However, had we found that all global anomalies from the continuum are
wiped out by the  Brillouin-Wigner phase convention there would have
been a reason to worry about it, or else, we would have been forced to
conclude that global gauge anomalies found in the continuum are not
to be taken seriously. 


\smallskip
\leftline{\sl 4d. An SU(2) example.}
\smallskip
The true analogue of $d=1$, $U(1)$ is $d=3$, $SU(2)$ 
since $\pi_1 (U(1)) = Z$ and $\pi_3 (SU(2)) = Z$. The $d=3$, $U(1)$
analysis above only helped us understand how global anomalies work on
the lattice. Also, since the matrices involved are 
smaller for $U(1)$, numerical computations can be done faster.
Unlike for $U(1)$, for $SU(2)$ one can observe the global gauge
anomaly even in a background that has zero field strength
everywhere. Moreover all loops winding round the tori can
be made trivial too, so the configuration is just a gauge transform
of the trivial one, with all link variables set to the unit matrix.
In a regularization employing the doubling trick, 
the global anomaly would assign relative 
weight one (with some normalization) to this 
configuration (A) and relative weight minus one to a configuration (B) 
which is obtained by gauge transforming the trivial configuration
by an element of the nontrivial
homotopy class of maps from the three torus to $SU(2)$. 

An example of
such an element is found in [\costeluscher] where it appeared
in momentum space in the Feynman diagram computation of the
induced Chern-Simons action. It is trivial to take the
continuum expression, view it now in real space and discretize it
to put it on the lattice. 
Actually, $V_f$ for a single component Dirac fermion
provides the map with regular spin being viewed
as isospin and regular space-time 
being viewed as momentum space. Let us call
the image of $V_f$, with a trivial factor in what is now spin space 
included, $G$. 
The mass parameter $m_0$ in $G$ controls the approximate
smoothness of the configuration on the 
lattice and we pick $m_0 \equiv m_0^G =1$. 
$$
(G)_{xi\alpha,yj\beta} = \delta_{\alpha,\beta} \delta_{x,y} 
{{-m_0^G \delta_{ij} +\sum_\mu [1-\cos ({{2\pi}\over L} x_\mu )]
\delta_{ij} 
+i\sum_\mu \sigma_{\mu, ij} \sin ({{2\pi}\over L} x_\mu )}\over
{\sqrt {[-m_0^G + \sum_\mu [1-\cos ({{2\pi}\over L} x_\mu )]]^2 +
\sum_\mu [ \sin ({{2\pi}\over L} x_\mu )]^2 }}}. \eqno{(4.6)}$$

To check for the global anomaly within the
Brillouin-Wigner convention all we
need to do now is to compare $z^{\rm BW}$ for the two 
configurations, both gauge equivalent to a trivial background.
This simplifies the numerical work significantly.  
$$
{{z^{\rm BW} (B)} \over {z^{\rm BW} (A)}} = {{\det (V_f G + G V_f )}\over 
{| \det (V_f G + G V_f )|}} {1\over \det (V_f)} .\eqno{(4.7)}$$
The parameter $m_0$ in $V_f$ controls how many degrees of freedom
represent massless fermions [\plbfirst]. This number 
is small when $m_0 \equiv m_0^f$
is close to zero. 

For $m_0^G =1$, $m_0^f =.8$ and $L=8$ 
we obtained $ -1$ for the ratio in (4.6),
the desired result. If we decrease either of these 3 parameters
the sign switch can be made to disappear. The masses can be made
smaller without loosing the sign 
switch if $L$ is increased. In the continuum
limit $L$ is taken to $\infty$ at 
fixed masses and the sign switch will stay.
We therefore established that the global anomaly occurs on the trivial
orbit for a large but smooth gauge transformation. 

We checked the stability of the finding on the trivial gauge orbit.
First, we perturbed the gauge transformation $G$ by sizable, but
limited, random perturbations. The sign switch relative to A remained.
Then we made the perturbations completely random. This wiped out
the sign switch. Once the gauge transformation is random it
does not matter whether one multiplies by $G$ also, or not.
There is no sign switch relative to configuration A. 
However, if the gauge transformations were random but limited,
following it up by $G$ produced a sign switch, while without $G$
the sign was the same as for A. 
Limited and unlimited fluctuating
gauge transformation (but with no $G$-factor)
have been observed not to give sign switches 
also in the case studied in [\narnish]. 
There the $|+>$ states were not taken
at infinite mass, so we do not {\it have} 
to be in agreement, but we are. Of course, all
statements about configurations with a certain
amount of randomness in them are of a statistical
nature. 

Similarly to the $U(1)$ case, 
not all sign switches mandated
by continuum arguments occur on the lattice. 
It is difficult to guess what exactly happens to the theory
if one carries out gauge averaging with the
Brillouin-Wigner phase convention. The configurations
that are typical of the continuum seem to cancel out because
of the global anomaly, but something is left over. The
problem is similar to the question what happens to an anomalous
theory in four dimensions if regulated by the overlap and gauge
averaged with the Brillouin-Wigner phase choice. 
Since in three dimensions the gauge violation is 
somewhat simpler, it might be easier to discover the fate
of the anomalous theories of this paper. 
A necessary first step would be to investigate other orbits,
with non-trivial gauge invariant content. We would like to
find out what kind of effective
action for the gauge invariant background is induced
by the gauge transformations
left over after cancelations due to the continuum global gauge
anomaly have taken effect. 
Any understanding we would reach would have to be checked
against variations of the Brillouin-Wigner phase convention.
After all, the fate of an anomalous gauge theory might very
well be non-universal.

\bigskip
\centerline{\bf 5. Summary.}
\medskip
The overall inter-dimensional relationships we see in this paper
indicate an intrinsic consistency of the overlap formalism across
different dimensions. We have seen examples where the overlap reproduces
global continuum anomalies. We therefore hope that the overlap
needs not only perturbative, but also global anomaly cancelation to
produce acceptable continuum theories after gauge averaging. There are
several applications of our formalism we could think of. An interesting
application (albeit a costly one) we are proposing is to ${\cal N} =1$
supersymmetric gauge theories in five dimensions where one could search
for nontrivial fixed points [\fivedim].

\medskip

\bigskip
\centerline{\bf 6. Acknowledgments.}
\medskip

H. N. was supported in part by the DOE under grant \#
DE-FG05-96ER40559. H. N. would like to thank L. Alvarez-Gaume
for making the suggestion (almost two years ago) 
to investigate the overlap in odd dimensions. 
We thank R. Narayanan for communications regarding this
work and [\narnish]. 

\bigskip
\centerline{\bf 7. References.}
\medskip

\item{[\npblong]} R. Narayanan, H. 
Neuberger, Nucl. Phys. B 443 (1995) 305.
\item{[\plbfirst]} R. Narayanan, H. 
Neuberger, Phys. Lett. B 302 (1993) 62.
\item{[\narnish]} R. Narayanan, 
J. Nishimura, hep-th/9703109 (1997).
\item{[\kimura]} T. Kimura, Prog. Theor. Phys. 92 (1994) 693.
\item{[\redlich]} A. N. Redlich, Phys. Rev. D29 (1984) 2366.
\item{[\alvarezgaumemoore]} L. Alvarez-Gaume, S. Della Pietra, G. Moore,
Ann. of Phys. 163 (1985) 288.
\item{[\costeluscher]} A. Coste, M. L{\" u}scher, 
Nucl. Phys. B323 (1989) 631.
\item{[\ho]} H. So, Prog. Theor. Phys. 73 (1985) 528; 74 (1985) 585.
\item{[\niemiseme]} A. J. Niemi, G. W. Semenoff, 
Phys. Rev. Lett. 51 (1983) 2077.
\item{[\witt]} E. Witten, Phys. Lett. 117B (1982) 324.
\item{[\elitzur]} S. Elitzur, V. P. Nair, Nucl. Phys. B243 (1984) 205.
\item{[\klinghammer]} F. R. Klinkhamer, Phys. Lett. B256 (1990) 41.
\item{[\nish]} N. Maru,  J. Nishimura, het-th/9705152 (1997).
\item{[\kapsch]} D. B. Kaplan, M. Schmaltz, Phys. Lett. B368 (1996) 44.
\item{[\kap]} D. B. Kaplan, Phys. Lett. B288 (1992) 342.
\item{[\anomnn]} R. Narayanan, H. Neuberger, 
Nucl. Phys. B412 (1994) 574.
\item{[\anomDS]} S. Randjbar-Daemi and J. Strathdee, 
Phys. Lett. B348 (1995) 543; Nucl. Phys. B443 (1995) 386; 
Phys. Lett. B402 (1997) 134; Nucl. Phys. B461 (1996) 305. 
\item{[\phasechoiceDS]} S. Randjbar-Daemi and J. Strathdee,
Nucl. Phys. B466 (1996) 335. 
\item{[\curcivene]} G. Curci, G. Veneziano, Nucl. Phys. B 292 (1987) 555.
\item{[\huetnn]} P. Huet, R. Narayanan, H. Neuberger, Phys. Lett. B380 
(1996) 291.
\item{[\schwartz]} L. Brink, J. H. Schwartz, 
J. Scherk, Nucl. Phys. B 121 (1977) 77.
\item{[\jackreb]} R. Jackiw, C. Rebbi, Phys. Rev. D13 (1976) 3398.
\item{[\dunne]} G. Dunne, K. Lee, C. Lu, Phys. 
Rev. Lett. 78 (1997) 3434.
\item{[\deser]} S. Deser, L. Griguolo, D. Seminara, hep-th/9705052 (1997).
\item{[\fosco]} C. Fosco, G. L. Rossini, F. A. Schaposnik, hep-th/9705124
(1997).
\item{[\holgerintanom]} H. B. Nielsen, M. Ninomiya, 
Int. J. Mod. Phys. A6 (1991) 2913.
\item{[\minfty]} D. Boyanovsky, E. Dagotto, E. Fradkin, Nucl. Phys. B 285 
(1987) 340; Y. Shamir, Nucl. Phys. B406 (1993) 90.
\item{[\412]} R. Narayanan, H. Neuberger, Phys. Lett. B 393 (1997) 360;
Y. Kikukawa, R. Narayanan, H. Neuberger, Phys. Lett. B 399 (1997).
\item{[\num412]} Y. Kikukawa, R. Narayanan, H. Neuberger, 
hep-lat/9705006 (1997).
\item{[\polyakov ]} A.M. Polyakov, Mod. Phys. Lett. A3 (1988) 325.
\item{[\hsu]} S. D. H. Hsu, hep-th/9704149 (1997).
\item{[\nnprl]} R. Narayanan, H. Neuberger, 
Phys. Rev. Lett. 71 (1993) 3251.
\item{[\nnu1]} R. Narayanan, H. Neuberger, Nucl. Phys. B477 (1996) 521.
\item{[\fivedim]} N. Seiberg, Phys. Lett. B387 (1996) 513.

\vfill
\eject
\end